\documentclass[11pt,twoside]{article}

\usepackage{edwin}
\usepackage{epsf}
\usepackage{psfig}
\usepackage{lscape}

\begin{document}
\title{Finding Astronomical Communities Through Co-readership Analysis}   
\author{Edwin A. Henneken, Michael J. Kurtz, Guenther Eichhorn, Alberto Accomazzi, Carolyn S. Grant, Donna Thompson, Elizabeth Bohlen, Stephen S. Murray}
\affil{Harvard-Smithsonian Center for Astrophysics, 60 Garden Street, Cambridge, MA 02138}

\begin{abstract} 

Whenever a large group of people are engaged in an activity, communities will form. The nature of these communities depends on the relationship considered. In the group of people who regularly use scholarly literature, a relationship like ``person i and person j have cited the same paper'' might reveal communities of people working in a particular field. On this poster, we will investigate the relationship ``person i and person j have read the same paper''. Using the data logs of the NASA/Smithsonian Astrophysics Data System (ADS), we first determine the population that will participate by requiring that a user queries the ADS at a certain rate. Next, we apply the relationship to this population. The result of this will be an abstract ``relationship space'', which we will describe in terms of various ``representations''. Examples of such ``representations'' are the projection of co-read vectors onto Principal Components and the spectral density of the co-read network. We will show that the co-read relationship results in structure, we will describe this structure and we will provide a first attempt in the classification of this structure in terms of astronomical communities.

The ADS is funded by NASA Grant NNG06GG68G.
\end{abstract}

\section{Introduction}

It has been shown (\citet{Kurtz05}) that well-defined ``modes of readership'' exist in the way people use the NASA Astrophysics Data System (ADS). It seems reasonable to expect that reading behavior is not random, especially in a reader's field of interest. For example, by browsing the Internet in general and the ADS in particular, through interaction with colleagues and by attending conferences, a researcher shares and learns about papers of interest. This will result in various types of patterns. People who work in the same field are more likely to cite each other and are more likely to be co-authors, thus forming patterns or ``communities'' (\citet{Newman06}). Readership itself is a more noisy medium and one can wonder whether patterns will have enough signal to emerge from the noise of, for example, journal browsing. The most straightforward way to find readership patterns is to look at ``co-readership'': the relationship ``person i and person j have read the same paper''. In the ADS data logs, every user is uniquely identified with a ``cookie ID'', which makes the determination of co-readership statistics a trivial exercise. The relationship defines an abstract space that can be described in various ways, depending on the interpretation we attribute to the relationship. Each approach has its merits. We can interpret the relationship as defining a network of nodes (representing individual users) and vertices (representing co-readership), and look for patterns by describing the topology of the network. Alternatively, we could be looking at a point cloud in a multidimensional space. The {\it crucial element} in our analysis will be the {\it choice of population}. Obviously, only people who use the ADS regularly will contribute in a meaningful way. One-time users, for example coming in through Google, only contribute to noise in the relationship space. Removing this noisy component is the easy part. The difficult part is translating ``people who use the ADS regularly'' into a real criterion. Do we want only those people who steadily read N times per month? And, what time interval should we choose? And there is the matter of selecting the journals we will monitor. The choice of population will be defined in ``Data'' section. On this poster we present our preliminary results based on Principal Component Analysis of the data. We represent the data by projecting co-read vectors on principal components. If there are correlations, we will be able to (drastically) reduce the dimensionality of the ``relationship space'' and we will see structure in the point cloud. Additionally we can check whether proximity of points in this point cloud can be associated with, for example, subject matters. A different way of looking for community structure is by determining the spectral density of the co-read network. In what way does this spectral density deviate from the spectral density of an uncorrelated random graph? This poster shows a first attempt in determining whether using readership data for community detection finds meaningful results.

\section{Data}

The source of our data consists of the Astrophysics Data System usage logs. We log all types of access by our users. An access ``type'' is related to which type of information is viewed for an article. We define ``reads'' as the access events by users, where multiple information retrievals per log period for one article by one user is regarded as a single ``read''. To rule out incidental use (e.g. by one-time users coming in via an external search engine, such as Google), we have taken the subset of users who query the database between 10 and 100 times per month.

As time interval we have taken the entire year of 2005. During this period we will determine reads to articles in the following core astronomy journals: {\it the Astrophysical Journal} (including Letters, and Supplement), {\it the Astronomical Journal}, {\it Astronomy \& Astrophysics}, {\it Monthly Notices of the Royal Astronomical Society} and {\it Publications of the Astronomical Society of the Pacific}. The number of ADS users in 2005, with a number of reads per month between 10 and 100, is about 10,000 (Henneken, 2006). From this population we pick our sample of $N_{s}$ users.

\medskip

\section{Results}

{\bf Principal Component Analysis}

For the $N_s$ users, we first determine the {\it co-read matrix} $\cal R$ with elements $r_{kl}$ ($k,l \in \{1,...,N_s\}$), which equal the number of reads that users $k$ and $l$ have in common. These (sample) users have been determined by first finding the total number of users (population) in the data set. If this number is larger than $N_s$, the population users are sorted according to total number of reads and the sample users are taken to be the first $N_s$ users of this set. We keep a mapping between user index and cookie ID. This will allow us, later on, to associate a point in ``co-read space'' to a common topic of papers (if any). From the {\it co-read matrix} $\cal R$, we determine the {\it normalized co-read matrix} $\cal N$, by normalizing the co-reads for each user by their total number of reads. The matrix $\cal R$ is {\it symmetrical}, and $\cal N$ is asymmetrical. We will use $\cal N$ for our analysis.

Next, the eigenvectors for $\cal N$ are determined. Figure 1 (left) shows the result for $N_s = 3000$ and $N_s = 4000$. The rank number refers to the rank number of the eigenvectors $\vec{e}_i$ ($i \in {1,...,N_s}$), where $\vec{e}_1$ has the largest eigenvalue ($\epsilon_1$). The inset shows the results for $\vec{e}_1$ through $\vec{e}_{20}$. Figure 1 (left) shows results for two values of $N_s$. Runs with other values for $N_s$ indicate that an increasing value of $N_s$ results in a higher eigenvalue for $\vec{e}_1$. Is our normalized co-read network an example of a scale-free network? Figure 1 (right) shows the relationship between $\epsilon_1$ and $N_s$, and the results show that $\epsilon_1 \propto N_s^\alpha$. The data result in a value for $\alpha$ of $0.1344$.

\begin{figure}[!ht]
  \plotone{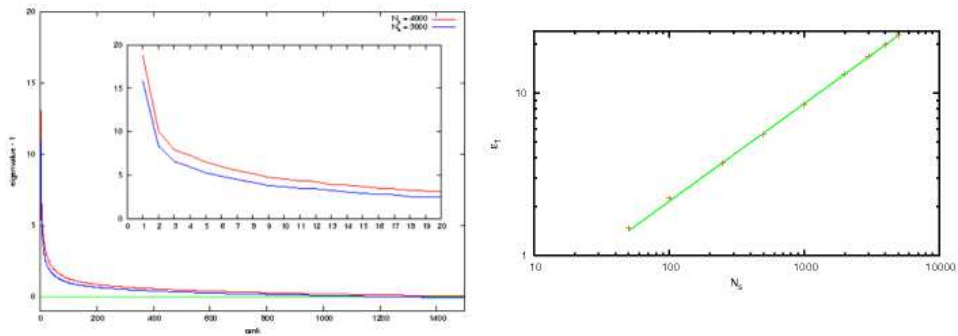}
  \caption{Left: eigenvalues for $N_s=3000$ (blue) and $N_s=4000$ (red). The inset shows a blow-up for eigenvectors $\vec{e}_1$ through $\vec{e}_{12}$. right: largest eigenvalue $\epsilon_1$ as a function of sample size $N_s$.}
\end{figure}

For the case $N_s=4000$, we project the co-read vectors onto the first three eigenvectors. The results are shown in figure 2. As already suggested by the distribution of eigenvalues, the presence of structure is obvious. Figure 3 shows a 3-dimensional representation of the projection of the co-read vectors onto the orthonormal basis spanned by $\vec{e}_1$ thru $\vec{e}_3$. The question now arises: can we associate this structure with ``communities''? In other words, can we relate proximity of points in figure 3 to, for example, a field of research? Every point in this 3-dimensional space can be traced back to an individual user. Therefore, every point in this space, has a set of papers associated with it. So, if we pick a point $\cal P$ in this space and take all points within a sphere around this point, we end up with a set of papers. As an example, we take point $(-0.05,0.2,0.03)$ (in figure 3), and choose the radius of the sphere to be $0.05$. We find 118 papers in this region. A reasonable, additional filter is citation count. The most cited papers in this set of papers are expected to be indicative of a common subject for these papers (if any). Looking at the set with 10 citations or more we find that most of these papers are about high-energy astrophysics, in particular about $\gamma$-ray astronomy, COMPTEL. Doing the same for point $(0.7,0.2,0.01)$, with a search radius of $0.01$, results in finding papers on SDSS, WMAP and galaxy classification.

\begin{figure}[!ht]
  \plotone{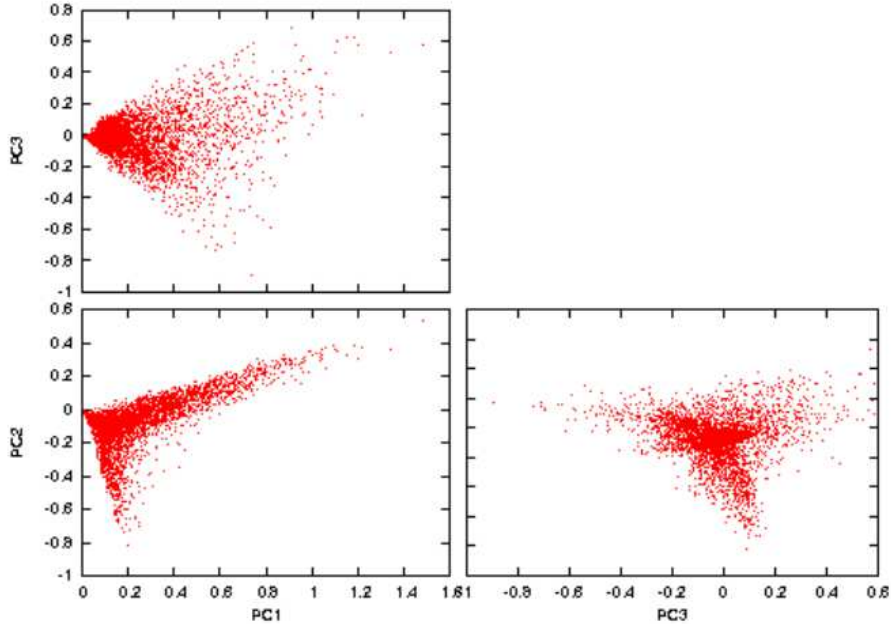}
  \caption{Projection of co-read vectors on principal components ($N_s=4000$). Top left: projection onto PC1 and PC3. Bottom left: projection onto PC1 and PC2. Bottom right: projection onto PC2 and PC3.}
\end{figure}

\begin{figure}[!ht]
  \plotone{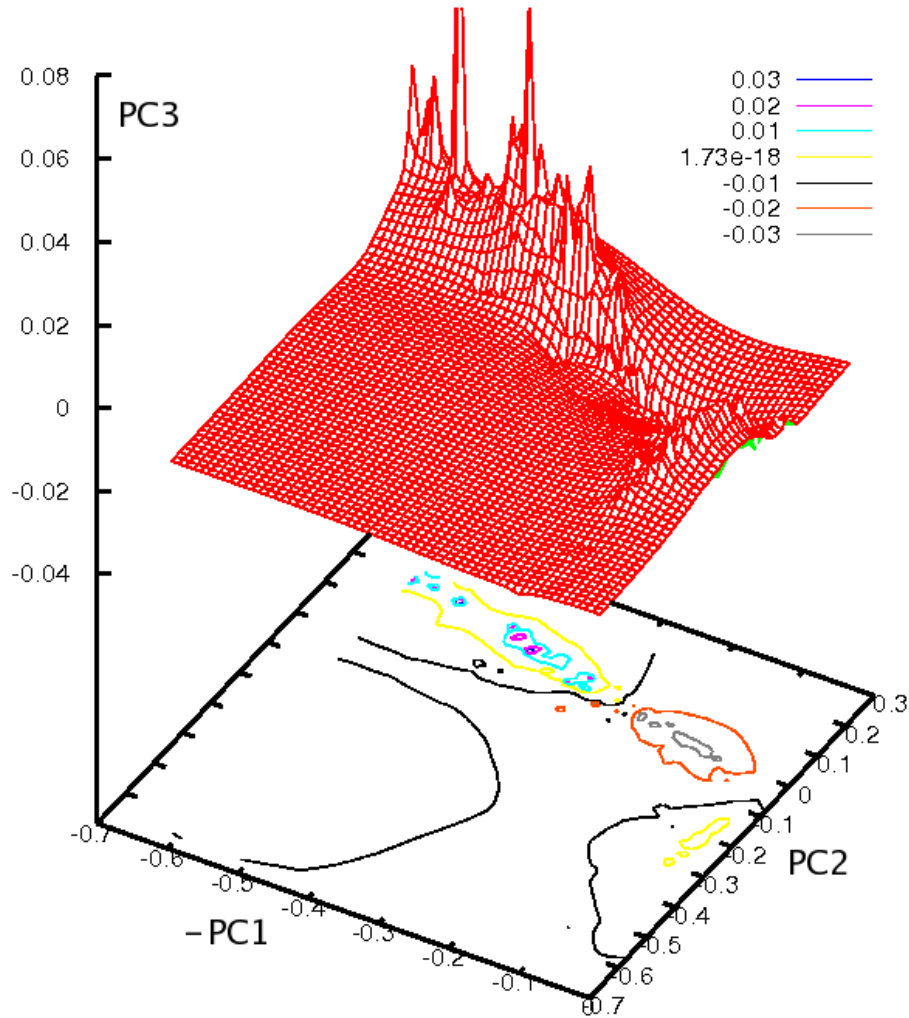}
  \caption{3-dimensional view of projection of co-read vectors on first three principal components}
\end{figure}

\smallskip
{\bf Spectral Density}

The spectral density $\rho$ of a graph is the density of the eigenvalues of its adjacency matrix. Spectral density measures the density of surrounding eigenvalues at each eigenvalue and serves as an especially useful metric of global graph topology (\citet{Farkas02}). Figure 4 shows the spectral density for our case of $N_s=4000$. In addition to the spectrum, the quantity $R$ (\citet{Farkas01}), defined as $(\epsilon_1-\epsilon_2)/(\epsilon_2-\epsilon_{N_s})$, also characterizes the type of network. It measures the distance of the first eigenvalue from the main part of the spectrum, normalized by the extension of the main part. Both the spectrum and shape of $R$ are consistent with a {\it scale-free network}: $\epsilon_1$ and the rest of the spectrum are well separated, and $R$ decays as a power law function of $N_s$. The value for $N_s=50$ for $R$ is smaller than what one would expect for a scale-free network, but this might be due to the small size. The results in figure 1 (right panel) are totally consistent down to $N_s=50$, though.

\begin{figure}[!ht]
  \plotone{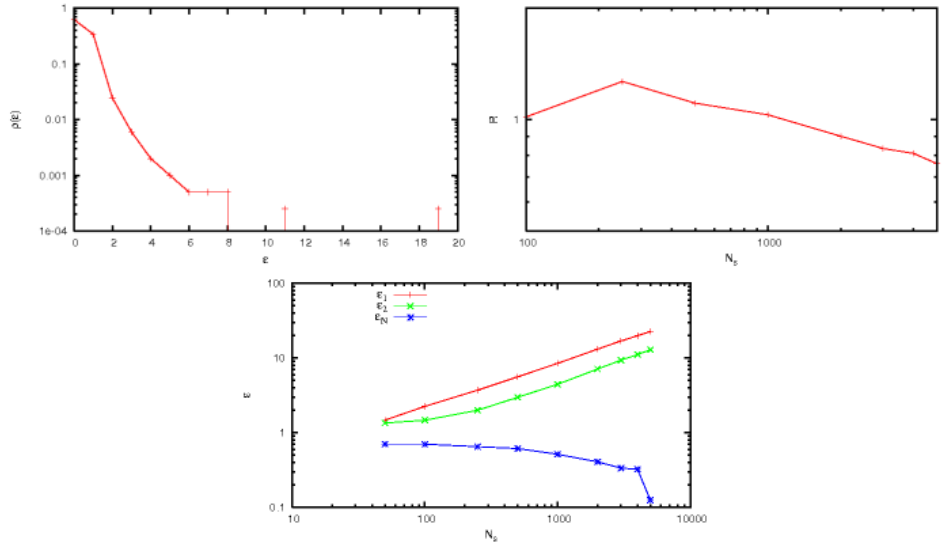}
  \caption{Top left: spectral density of eigenvalues. top right: the quantity R as a function of sample size $N_s$. bottom: eigenvalues as function of $N_s$}
\end{figure}

\section{Discussion}

The first results indicate that co-readership networks, at least within the population used here, are strongly correlated. Furthermore, proximity in the space spanned by the eigenvectors with the three largest eigenvalues (figures 2 and 3), seems to have real meaning in terms of subject areas of papers.

Metrics based on co-readership data (as shown in figures 1 and 4) are consistent with the characteristics of a scale-free random graph. A {\it scale-free random graph} consists of a growing set of vertices and edges, where the location of the new edges is determined by a {\it preferential attachment rule}. This seems like a reasonable process to describe the dynamics of a co-readership network.

We find that the largest eigenvalue $\epsilon_1$ grows like $N_s^\alpha$, with $\alpha=0.1344$. This differs from the characteristic $\alpha=0.25$ (\citet{Goh01}) which one usually sees for large enough scale-free systems. This probably results from the fact that we used the (asymmetric) normalized co-readership matrix for our analysis, which is more like a Laplacian matrix than an adjacency matrix. Another influence is the fact that we disregarded contributions from users with less than 10 reads per month. Thus we exclude nodes that contribute heavily to the low-degree component.

The results presented here are the preliminary outcome of a first exploration of co-readership data. We wanted to establish whether it is possible to detect meaningful structure within the intrinsically noisy readership data. The results presented here indicate that this is indeed the case. In future work, we need to establish the influence of the various parameters that determined the population used in our analysis.


\end{document}